\documentclass[nofootinbib,aps,pra,twocolumn,showkeys,showpacs,preprintnumbers,floatfix]{revtex4-1}

\usepackage{dynlearn}
\usepackage{listings}

\usepackage{color}
\usepackage{braket}

\definecolor{dkgreen}{rgb}{0,0.6,0}
\definecolor{gray}{rgb}{0.5,0.5,0.5}
\definecolor{mauve}{rgb}{0.58,0,0.82}

\begin{document}

\def\ourTitle{%
Optimizing Quantum Models of Classical Channels:\\
The reverse Holevo problem
}

\def\ourAbstract{%
Given a classical channel---a stochastic map from inputs to outputs---the input
can often be transformed to an intermediate variable that is informationally
smaller than the input. The new channel accurately simulates the original but
at a smaller transmission rate. Here, we examine this procedure when the
intermediate variable is a quantum state. We determine when and how well
quantum simulations of classical channels may improve upon the minimal rates of
classical simulation. This inverts Holevo's original question of quantifying
the capacity of quantum channels with classical resources. We also show that
this problem is equivalent to another, involving the local generation of a
distribution from common entanglement.
}

\def\ourKeywords{%
quantum information, classical channel, complexity.
}

\hypersetup{
  pdfauthor={},
  pdftitle={\ourTitle},
  pdfsubject={\ourAbstract},
  pdfkeywords={\ourKeywords},
  pdfproducer={},
  pdfcreator={}
}
\author{Samuel P. Loomis}
\email{sloomis@ucdavis.edu}

\author{John R. Mahoney}
\email{jrmahoney@ucdavis.edu}

\author{Cina Aghamohammadi}
\email{caghamohammadi@ucdavis.edu}

\author{James P. Crutchfield}
\email{chaos@ucdavis.edu}
\affiliation{Complexity Sciences Center and Physics Department,
University of California at Davis, One Shields Avenue, Davis, CA 95616}

\date{\today}
\bibliographystyle{unsrt}

%%%%%%%%%%%%%%%%%%%%%%%%%%%%%%%%%%%%%%%%%%%%%%%%%%%%%%%%%%%%%%%%%%%%%%%%%%%%%%%
% The paper content

\title{\ourTitle}

\begin{abstract}
\ourAbstract
\end{abstract}



\preprint{\arxiv{1709.08101 [quant-ph]}}

\title{\ourTitle}
\date{\today}
\maketitle
% \tableofcontents

\setstretch{1.1}

%\listoffixmes

%\tableofcontents

%\fxnote{For each lemma, note (at least in the tex) how strong it is with respect to pure vs mixed, PVM vs POVM, etc.}

\section{Introduction}
\label{sec:introduction}
One speaks of a quantum advantage when a computational task is performed more
efficiently (in memory, time, or both) using quantum mechanical hardware than
classical hardware. Quantum advantages appear in the simulation of a variety of
classical systems \cite{Meye02aa}: thermal states \cite{Yung10aa}, fluid flows
\cite{Yepe01aa,Yepe99a}, electromagnetic fields \cite{Sinh10a}, diffusion
processes \cite{Yepe01ab,Berm02aa}, Burger's equation \cite{Yepe02aa}, and
molecular dynamics \cite{Harr10aa}. Quantum advantage also has been found in
more mathematical contexts. The most well-known problems include the
factorization of prime numbers (Shor's integer factoring algorithm
\cite{Shor99aa}), database search (Grover's algorithm \cite{Grov96aa}), and the
efficient solution of linear systems \cite{Harr09aa}. 

A recent but rich area of study is the quantum advantage for simulating
classical stochastic processes. By stochastic process, we mean a source that
probabilistically generates a sequence of symbols $x_0 x_1 \dots x_t$. When the
probability of each new symbol $x_t$ depends on the previous symbols, the
process is said to have memory. Computational mechanics has previously studied
the memory resources required for simulating and predicting stochastic
processes using classical hardware \cite{Crut88a,Shal98a,Crut12a}. Quantum
computational mechanics now proposes to use sequential measurements of a
quantum system as a more resource-efficient means of simulating stochastic
processes \cite{Gu12a,Tan14a,Maho16a,Riec16a,Bind17a}.

Recent efforts showed that many important results on optimal simulation and
prediction do not carry over from the classical to the quantum domain. For
instance, it is a common fact of resource theories that a given resource (say,
memory) may have various inequivalent quantifications depending on the specific
task at hand (say, asymptotic rates versus single-shot requirements).  In
classical computational mechanics, in contrast, a stochastic process has a
uniquely optimal predictive model---the \eM. It is minimal according to all
quantifications of memory. In this sense, the \eM \emph{strongly minimizes}
memory \cite{Loom18a}. However, this no longer holds for quantum models, where
inequivalent quantifications have inequivalent minima. For example, a model
that is optimal for single-shot implementations is not the optimal model for
the asymptotic rate. We call this \emph{weak optimization}
\cite{Loom18a,Liu18a}.

In a memoryful stochastic process, the future is dependent on the past through
what is mathematically considered a probabilistic channel. This motivates the
exploration of simulating classical channels with quantum resources, which is a
fundamental question that sheds light on the more complicated problem of
simulating stochastic processes. This question is the main motivator of the
following development, although we call-out ancillary results along the way.

Simulating a classical channel with quantum resources is in
many ways the inverse of much previous work on channels in quantum information.
There, the focus is often on the resources (such as entanglement) required to
simulate a fully quantum channel \cite{Benn99a,Benn02a,Benn14a}. It also
often addresses the capacity of a quantum channel to transmit classical
information, first considered by Holevo \cite{Hole73a,Schu97a,Hole98a,Hole98b}.
Rather than using classical capacities as a means to study the properties of
quantum channels, the following uses quantum resources as a means to study the
properties of purely classical channels. Thus, in the spirit of the ``reverse
Shannon theorem'' \cite{Benn02a}, we consider this a \emph{reverse Holevo
problem}.

Our results also draw from the literature of common information
\cite{Wyne75a,Kuma14a}, which concerns simulating channels with intermediate
classical variables. We adapt those results to the quantum domain by making the
intermediate variable quantal.

The following (second) section establishes notation and definitions for our
development. The third presents our three main results for the reverse
Holevo problem:\\[-20pt]
\begin{enumerate}
      \setlength{\topsep}{-25pt}
      \setlength{\itemsep}{-5pt}
      \setlength{\parsep}{-5pt}
      \setlength{\labelwidth}{5pt}
      \setlength{\itemindent}{-5pt}
\item For every channel, there are quantum models that reduce memory costs
	across all quantifications of memory (or, in rare cases, do at least as
	well as the original channel).
\item However, there is generally not a single quantum model that is minimal
	with respect to all quantifications of memory. This has important
	implications for single-shot channel simulation.
\item We demonstrate a lower bound for the asymptotic cost of modeling a
	channel with quantum resources.
\end{enumerate}
We also present a mathematical correspondence between the reverse Holevo
problem and another that we call \emph{common entanglement}. The latter asks
for the entanglement cost of generating a bivariate classical distribution
using only entanglement and local operations (without communication).
This correspondence sets the stage for further developments.

%%%%%%%%%%%%%%%%%%%%

\section{Notation and definitions}
\label{sec:notation}

This section adapts results from computational mechanics, as well as from the
theory of common information, to describe Shannon communication channels.

Computational mechanics is a subfield of statistical mechanics that addresses
the information-theoretic and energetic costs of simulating and predicting
stochastic processes \cite{Crut12a,Boyd17a}. In that setting, a \emph{stochastic
process} is defined by an alphabet $\mathcal{A}$ and a probability measure
$\mathbb{P}\left(x_{-\infty:\infty}\right)$ over all bi-infinite words
$x_{-\infty:\infty}=\dots x_{-1} x_0 x_{1} \dots$ such that $x_t\in\mathcal{A}$
for all $t\in\mathbb{Z}$. 

Random variables are denoted by capital Latin letters $X$, $Y$, and so on, and
their outcomes by $x\in\mathcal{X}$, $y\in\mathcal{Y}$, and so on. The
calligraphic letters $\mathcal{X},\mathcal{Y}$, and so on represent finite
sets.

Classical distributions are denoted by $\mathbb{P}$. For example, the
probability of a random variable $X$ taking value $x\in\mathcal{X}$ is given by
$\mathbb{P}(X=x)$ if the random variable needs to be specified and is shortened
to $\mathbb{P}(x)$ if it can be inferred from context. We say $Y=f(X)$ for a
function $f:\mathcal{X}\rightarrow \mathcal{Y}$ if $\mathbb{P}\left(Y=y\right)
= \sum_{x:f(x)=y} \mathbb{P}\left(X=x\right)$.

As a measure of its information content, the uncertainty of a random variable
is measured by its \emph{R{\'e}nyi entropies}:
\begin{align*}
H_\alpha\left(X\right)
  := \frac{1}{1-\alpha}
  \log \left(\sum_{x\in\mathcal{X}}\mathbb{P}(x)^\alpha\right)
  ~,
\end{align*}
for $\alpha\in (0,1)$ and $\alpha\geq 1$. In the limits $\alpha\rightarrow 0$
and $\alpha\rightarrow 1$, we recover the \emph{max-entropy}
$H_0\left(X\right):=\log \left|\mathcal{X}\right|$ (useful for costs in
single-shot or zero-error situations) and the \emph{Shannon entropy} $H(X) :=
-\sum_{x} \mathbb{P}(x)\log \mathbb{P}(x)$, respectively. An important property
of the R{\'e}nyi entropies is that they are monotonically decreasing under the
application of a deterministic function. That is, if $Y=f(X)$, then $H(Y)\leq
H(X)$, with equality only when $f$ is bijective.

\subsection{Classical channels}

When discussing channels, we consider an input space $\mathcal{X}$ and output
space $\mathcal{Y}$, both (for convenience) assumed to be discrete. On these
one may define random variables $X$ and $Y$, respectively. A \emph{channel}
$\mathcal{C}$ is then a conditional probability function
$\mathbb{P}_{\mathcal{C}}\left(Y=y\middle|X=x\right)$, which for each outcome
$x\in\mathcal{X}$ in the input space defines a probability distribution over
the output space $\mathcal{Y}$. We write such a channel as
$\mathcal{C}:\mathcal{X}\rightsquigarrow \mathcal{Y}$, distinguishing it from a
function via the squiggly arrow.

When it comes to information theory, a channel's associated costs and resources
depend on the input distribution $\mathbb{P}\left(X=x\right)$. One may vary
these inputs to determine maximum capacities or minimum costs, for example.
When speaking of a channel $\mathcal{C}$, though, for completeness we assume
that an input distribution has been defined and so $XY$ forms a bivariate
random variable. $XY$ is defined on the space $\mathcal{X}\times\mathcal{Y}$
with probability distribution $\mathbb{P}\left(X=x,Y=y\right) =
\mathbb{P}_{\mathcal{C}}\left(Y=y\middle|X=x\right)\mathbb{P}\left(X=x\right)$.
This assumption does not affect the generality of our results.

For bivariate random variables, the Shannon entropy can be used to define the
\emph{conditional entropies} and \emph{mutual information}:
\begin{align*}
H\left(X\middle| Y\right) &:= H\left(XY\right) - H\left(Y\right) ~,\\
H\left(X\middle| Y\right) &:= H\left(XY\right) - H\left(Y\right) ~,~\text{and}\\
I\left(X: Y\right) &:= H\left(X\right) + H\left(Y\right) - H\left(XY\right)
  ~.
\end{align*}
These represent the uncertainty remaining in $X$ when $Y$ is known,
the uncertainty in $Y$ when $X$ is known, and the information shared
between $X$ and $Y$, respectively. 

We wish to {\em simulate} a particular channel $\mathcal{C}$. That is, we want
(i) a random variable $Z$ defined on an intermediate space $\mathcal{Z}$, (ii)
an encoding channel $\mathcal{E}:\mathcal{X}\rightarrow \mathcal{Z}$ with
conditional distribution $\mathbb{P}_\mathcal{E}\left(Z=z\middle|X=x\right)$,
and (iii) a decoding channel $\mathcal{D}:\mathcal{Z}\rightsquigarrow
\mathcal{Y}$ with conditional distribution
$\mathbb{P}_\mathcal{D}\left(Y=y\middle|Z=z\right)$ such that:
\begin{align*}
\mathbb{P}_{\mathcal{C}}\left(y\middle|x\right)
  = \sum_{z\in\mathcal{Z}} \mathbb{P}_\mathcal{D}\left(y\middle|z\right)
  \mathbb{P}_\mathcal{E}\left(z\middle|x\right) 
 ~.
\end{align*}
We write this composition rule as $\mathcal{C} = \mathcal{D}\circ \mathcal{E}$. 
We call the triplet $M=\left(\mathcal{Z},\mathcal{E},\mathcal{D}\right)$
a {\em model} of channel $\mathcal{C}$.

Additionally, $X-Z-Y$ form a Markov chain. The \emph{data processing
inequality} tells us that $H(Z)\geq I\left(X:Z\right)\geq I\left(X:Y\right)$.
This puts a constraint on models that simulate a given channel $\mathcal{C}$.

It is helpful to consider a particular subset of models called {\em
factorizations} or, simply, \emph{factors}. These are triples
$F=\left(\mathcal{Z},f,\mathcal{D}\right)$, where the encoder is a
deterministic function $f:\mathcal{X}\rightarrow \mathcal{Z}$.

\begin{figure}
\includegraphics[width=0.75\linewidth]{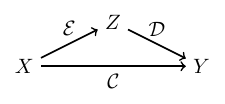}
\caption{Information flow when simulating a channel: When $(\mathcal{Z},
	\mathcal{E},\mathcal{D})$ is a model, all paths in the diagram from $X$ to
	$Y$ represent the same channel---the diagram commutes. When
	$\mathcal{E}=f$ is a deterministic function, the model is a
	factor.
}
\label{fig:model_diagram}
\end{figure}

Factors and models may be visualized by the commuting diagram of Fig.
\ref{fig:model_diagram}.

What is the use of a model or a factor? Their use arises when considering
channels as a resource. If we do not have access to a physical system that
directly implements channel $\mathcal{C}$, we may resort to systems that
instead implement channels $\mathcal{E}$ and $\mathcal{D}$ in a model
$M=\left(\mathcal{Z},\mathcal{E},\mathcal{D}\right)$ of $\mathcal{C}$.  The
more information required to store the intermediate variable $Z$, the more
``costly'' the model.

Models may be compared in quantifiable terms---through measurable features,
such as the R{\'e}nyi entropies of the intermediate variable $Z$---or they may
be compared operationally. On the quantifiable side, one may pick a particular
value of $\alpha$ and ask which model has the larger $H_\alpha\left(Z\right)$.
Given two models $M_1=(\mathcal{Z}_1,\mathcal{E}_1,\mathcal{D}_1)$ and
$M_2=(\mathcal{Z}_2,\mathcal{E}_2,\mathcal{D}_2)$ one often finds that $M_1$ is
advantageous with respect to some $\alpha$---$H_\alpha\left(Z_1\right)\leq
H_\alpha\left(Z_2\right)$---while $M_2$ is advantageous with respect to
others---$H_\alpha\left(Z_2\right)\leq H_\alpha\left(Z_1\right)$.

Sometimes, however, the advantage of $M_1$ over $M_2$ is unilateral. Given two
models $M_1$ and $M_2$ of a channel $\mathcal{C}$, we say that $M_1$ is
\emph{strongly advantageous} with respect to $M_2$ if, for all $\alpha\geq 0$,
$H_\alpha\left(Z_1\right)\leq H_\alpha\left(Z_2\right)$.

This differs from being \emph{strictly} advantageous, which implies for at
least one $\alpha$ we have $H_\alpha\left(Z_1\right)< H_\alpha\left(Z_2\right)$
without equality.

\begin{figure}
\includegraphics[width=0.75\linewidth]{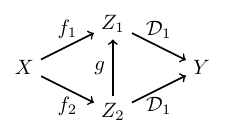}
\caption{Channel sufficiency: $F_2=(\mathcal{Z}_2,f_2,\mathcal{D}_2)$ is
	sufficient for $F_1=(\mathcal{Z}_1,f_1,\mathcal{D}_1)$ if, for some
	deterministic $g$, this diagram commutes.
}
\label{fig:factor_diagram}
\end{figure}

In contrast, the operational approach asks whether one model of a channel
$\mathcal{C}$ can effectively substitute for another. For factorizations, there
is a particularly useful operational comparison: given factorizations
$F_1=(\mathcal{Z}_1,f_1,\mathcal{D}_1)$ and
$F_2=(\mathcal{Z}_2,f_2,\mathcal{D}_2)$ of channel $\mathcal{C}$, we say that
$F_2$ is \emph{sufficient} for $F_1$ if there is a deterministic\footnote{The
mapping $g$ need not be explicitly deterministic, but if $f_1$ and $f_2$ are
deterministic, then this will require $g$ to be so as well.} function
$g:\mathcal{Z}_2\rightarrow \mathcal{Z}_1$ such that Fig.
\ref{fig:factor_diagram}'s diagram commutes.

The idea behind this comparison is that factorization $F_2$ itself can be
factorized to yield the model $F_1$. In effect, $F_2$ already uses all the
memory required to implement $F_1$, as well as some additional overhead. The
takeaway, then, is that we might as well use $F_1$, as $F_2$ costs at least as
much.

As a consequence of the concavity of R{\'e}nyi entropies, the statement ``$F_2$
is sufficient for $F_1$'' necessarily implies that $F_1$ is strongly
advantageous with respect to $F_2$. If $F_2$ is sufficient for $F_1$ and the
function $g:\mathcal{Z}_2\rightarrow \mathcal{Z}_1$ is not bijective, then
$F_1$ is strictly advantageous with respect to $F_2$.

Our focus on factorizations pays off with the following result from
computational mechanics: For each channel $\mathcal{C}$, there exists a
\emph{causal factorization} $F_\mathcal{C}$, unique up to isomorphism, for
which all other factorizations $F$ of $\mathcal{C}$ are sufficient
\cite{Shal98a}. Consequently, $F_\mathcal{C}$ is strongly minimal with respect
to all other factorizations. The rule that
$F_\mathcal{C}(x_1)=F_\mathcal{C}(x_2)$ if and only if
$\mathbb{P}\left(Y=y\middle|X=x_1\right)=\mathbb{P}\left(Y=y\middle|X=x_2\right)$
defines $F_\mathcal{C}$ up to an isomorphism \cite{Crut88a}. (As a consequence,
the image of the factorization $Z_\mathcal{C}$ is the minimal sufficient
statistic of $X$ with respect to $Y$.)

Notice, that the data processing inequality provides the lower bound
$H\left(Z\right)\geq I\left(X:Y\right)$ for all models and factorizations of
$\mathcal{C}$. This lower bound is generally not achievable with
factorizations. 

One might conjecture that the nonachievability of $I(X:Y)$ is a consequence of
constraining our models to factorizations. However, the problem extends
further. The more general problem of optimal \emph{models} (in the broader,
stochastic sense defined above) falls under the purview of common information
theory \cite{Wyne75a,Kuma14a}. \emph{Common information} refers to the problem
of determining when a bivariate distribution $XY$ may be jointly generated from
a common variable $Z$ by local stochastic maps. That is, if $Z$ is distributed
according to $\mathbb{P}\left(Z=z\right)$, then $XY$ is distributed according
to:
\begin{align*}
\mathbb{P}\left(x,y\right)
  = \sum_{z\in\mathcal{Z}} \mathbb{P}_A\left(x\middle|z\right)
  \mathbb{P}_B\left(y\middle| z\right) \mathbb{P}\left( z\right) 
\end{align*}
for channels $\mathbb{P}_A$ and $\mathbb{P}_B$. Any such joint generation
scheme can be rewritten as a model of the channel $\mathcal{C}$ from
$\mathcal{X}$ to $\mathcal{Y}$. In this way, results in the field of common
information carry over to models of channels.

In particular, Wyner introduced a measure of common information that relates
the asymptotic problem of modeling to the single-shot problem \cite{Wyne75a}.
In the asymptotic regime, we can discuss a more general type of model. A
$(n,M,\epsilon)$-\emph{model} of $\mathcal{C}$ is a pair of channels
$\mathcal{E}:X\rightarrow Z$ and $\mathcal{D}:Z\rightarrow Y$ such that
$\mathcal{E}\left(X\right)$ is uniform over $M$ outcomes and:
\begin{align*}
\left\| \widehat{\mathbb{P}}\left(x,y\right)
  - \mathbb{P}(x,y)\right\|_1<\epsilon
  ~,
\end{align*}
where $\widehat{\mathbb{P}}\left(x,y\right)= \mathbb{P}_{\mathcal{D}\circ\mathcal{E}}\left(y\middle|x\right) \mathbb{P}(x)$.

That is, $(n,M,\epsilon)$ is a model that approximates $n$ independent and
identically distributed copies of the channel $\mathcal{C}$ with entropy cost
$H(Z) = \log M$ and error $\epsilon$. Error here is measured by the
$\ell^1$-norm $\left\|\cdot\right\|_1$. Wyner's result, strengthened by Winter
and Ahlswede \cite{Ahls02a,Wint05a}, is that for any $\epsilon,\delta>0$ there
exists a sufficiently large $n$ such that an $(n,M,\epsilon)$ model is Markov
with $\frac{1}{n}\log M \leq C\left(X:Y\right) + \delta$, where:
\begin{align}
C\left(X:Y\right) = \min_{X-Z-Y} I\left(XY:Z\right)
\label{eq:wyner}
\end{align}
and the minimum is taken over all $Z$ such that $X-Z-Y$. Conversely, for any
$\delta>0$, there exists a $\Delta>0$ such that all $(n,M,\epsilon)$-models
with $\frac{1}{n}\log M\leq C(X:Y) - \delta$ have error at least $\epsilon
>\Delta$. In short, $C\left(X:Y\right)$ is the minimal asymptotic cost rate at
which a channel can be approximately simulated.

Usually, $C(X:Y) > I(X:Y)$. And so, as in the case of factorizations,
stochastic models of channels cannot reduce the memory cost all the way down to
the mutual information.

The concepts just reviewed are various and diverse. After introducing
equivalent notions for quantum models of channels, we will provide two examples
that help elucidate them.

%%%%%%
\subsection{Quantum models of channels}

The notion of a classical model readily generalizes to one of a \emph{quantum
model} for a classical channel. For clarity, however, we start by defining the
important concepts from quantum information.

Instead of random variables, we have quantum states---positive operators $\rho$
on a Hilbert space $\mathcal{H}$ such that $\mathrm{Tr}\left(\rho\right)=1$. We
assume $\mathcal{H}$ is finite-dimensional with dimension $d$, so that $\rho$
can be decomposed in the form:
\begin{align*}
\rho = \sum_{i=1}^d \lambda_i \ketbra{i}
  ~,
\end{align*}
for an orthogonal basis $\left\{\ket{i}\right\}$ and coefficients $\lambda_i>0$
satisfying $\sum_i \lambda_i=1$. The coefficients can be thought of as a probability
distribution. Consequently, we can generalize the R{\'e}nyi entropies to the
quantum setting:
\begin{align*}
S_\alpha\left(\rho\right) := & \frac{1}{1-\alpha}
  \log \left(\sum_{i=1}^d \lambda_i^\alpha\right) \\
  = & \frac{1}{1-\alpha} \log\mathrm{Tr} \rho^\alpha
  ~.
\end{align*}
In the limits $\alpha\rightarrow 0$ and $\alpha\rightarrow 1$, we obtain the
\emph{max-entropy} $S_0\left(\rho\right):=\log \dim\mathcal{H}$ (useful for
costs in single-shot or zero-error situations) and the \emph{von Neumann
entropy} $S(\rho) := -\sum_{i=1}^d \lambda_i \log \lambda_i =
-\mathrm{Tr}\left(\rho \log \rho\right)$, respectively.

The interface between the quantum and classical world relies on two classes of
operation. The first is an \emph{ensemble}, represented here as a function
$\mathcal{E}:\mathcal{X}\rightarrow \mathcal{S}\left(\mathcal{H}\right)$ from a
finite set $\mathcal{X}$ to the set $\mathcal{S}\left(\mathcal{H}\right)$ of
states on $\mathcal{H}$. It denotes a preparation of a quantum system using
some initial classical information, stored in a random variable $X$. A
\emph{pure-state ensemble} is one for which $\mathcal{E}\left(x\right) =
\ketbra{\psi_x}$ for some (not necessarily orthogonal) vectors
$\left\{\ket{\psi_x}\right\}$.

The second operation is \emph{measurement}, represented as a stochastic map
$\mathcal{M}:\mathcal{S}\left(\mathcal{H}\right)\rightarrow \mathcal{Y}$ from
the set of states to a finite set $\mathcal{Y}$, with the constraint that
$\mathbb{P}_\mathcal{M}\left(y\middle|\rho\right) = \mathrm{Tr}\left( M_y
\rho\right)$ for some set of positive operators $\left\{ M_y\right\}$
satisfying $\sum_y M_y = 1$. This is known as a \emph{positive-operator-valued
measure} or POVM. We have a \emph{projector-valued measure} or PVM when $M_y =
\ketbra{y}$ for some orthogonal basis $\left\{\ket{y}\right\}$.

Ensembles and measurements may be composed to create a classical channel. We
say $\mathcal{M}\circ \mathcal{E} = \mathcal{C}$ if:
\begin{align*}
\mathbb{P}_\mathcal{C}\left(y\middle|x\right) = \mathrm{Tr}\left(M_y \mathcal{E}(x)\right)
  ~.
\end{align*}

We are now ready to define a \emph{quantum model of a channel} $\mathcal{C}$.
This is the triplet $\mathcal{Q}=\{\mathcal{H},\mathcal{E},\mathcal{M}\}$
containing a Hilbert space $\mathcal{H}$, an ensemble
$\mathcal{E}:\mathcal{X}\rightarrow \mathcal{B}(\mathcal{H})$, and a POVM
$\mathcal{M}:\mathcal{B}(\mathcal{H})\rightsquigarrow \mathcal{Y}$, such that
$\mathcal{M}\circ \mathcal{E} = \mathcal{C}$.

Classically, we found it useful to specify \emph{factors} of channels, which
replaced the stochastic encoding map with a deterministic one. Here, the
closest analogue is a model
$\mathcal{Q}=\{\mathcal{H},\mathcal{E},\mathcal{M}\}$ where $\mathcal{E}$ is a
pure-state ensemble. We call this a \emph{pure-state model}.

However, each quantum model
$\mathcal{Q}=\{\mathcal{H},\mathcal{E},\mathcal{M}\}$ also \emph{induces} a
classical factorization.  The induced factorization of a model $\mathcal{Q}$ is
the classical factorization
$F_\mathcal{Q}=\left(\mathcal{Z}_\mathcal{Q},f_{\mathcal{Q}},\mathcal{D}_\mathcal{Q}\right)$
of $\mathcal{C}$ such that $f_{\mathcal{Q}}(x)=f_{\mathcal{Q}}(x')$ if and only
if $\rho_x=\rho_{x'}$. This factorization is unique up to isomorphisms. Note
that each induced factorization must be a \emph{refinement} of the causal
factorization $F_\mathcal{C}$; that is, $F_Q(x_1)=F_Q(x_2)$ implies
that $F_\mathcal{C}(x_1)=F_\mathcal{C}(x_2)$.

The notion of strongly advantageous carries over to the quantum domain. Let
$\mathcal{Q}_1=\{\mathcal{H}_1,\mathcal{E}_1,\mathcal{M}_1\}$ and
$\mathcal{Q}_2=\{\mathcal{H}_2,\mathcal{E}_2,\mathcal{M}_2\}$ be two quantum
models. For a given input $\mathbb{P}(x)$, let $\rho_1 = \sum_x \mathbb{P}(x)
\mathcal{E}_1(x)$ and $\rho_2 = \sum_x \mathbb{P}(x) \mathcal{E}_2(x)$
represent the model states. We say that $\mathcal{Q}_1$ is \emph{strongly
advantageous} with respect to $\mathcal{Q}_2$ if $S_\alpha(\rho_1) \leq
S_\alpha(\rho_2)$ for all $\alpha\geq 0$.

We can also generalize the asymptotic notion of models. An
$(n,M,\epsilon)$-quantum model is given by an $M$-dimensional space
$\mathcal{H}$, an ensemble $\mathcal{E}:\mathcal{X}^N\rightarrow
\mathcal{B}\left(\mathcal{H}\right)$, and
$\mathcal{M}=\left\{M_{y^n}\in\mathcal{B}\left(\mathcal{H}\right)
:y^n\in\mathcal{Y}^n\right\}$, such that $\rho = \sum_{x^n} \mathbb{P}({x^n})
\mathcal{E}(x^n)$ is uniform over $\mathcal{H}$ and:
\begin{align*}
\left\| \widehat{\mathbb{P}}\left(x,y\right) -
\mathbb{P}(x^n,y^n)\right\|_1<\epsilon
  ~,
\end{align*} 
defining $\widehat{\mathbb{P}}\left(x^n,y^n\right):=
\mathbb{P}_{\mathcal{M}\circ\mathcal{E}}\left(y^n\middle|x^n\right)
\mathbb{P}(x)$.

%%%%%%%
\subsection{Examples}

\begin{figure}
\includegraphics[width=\linewidth]{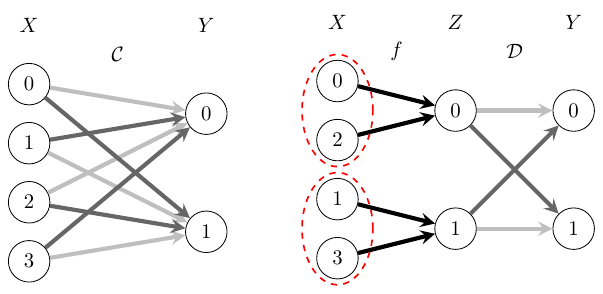}
\caption{Factorizing the Redundant Binary Symmetric Channel: (Left) Channel
	from $X$ to $Y$. (Right) Intermediate variable $Z$ cleaves channel into two
	steps---first deterministic, then stochastic. Arrow shade indicates
	conditional probability: black: $1$, dark gray: $1-p$, and light gray: $p$.
	Dashed ellipses show input equivalence classes. (Note re-ordering of
	$X = 1$ and $X = 2$.)
	}
\label{fig:RBSC_classical_factorization}
\end{figure}

The \emph{Redundant Binary Symmetric Channel} \cite{Ash90a} (RBSC) 
is a simple example that compactly illustrates the potential for
quantum advantage in simulating classical channels. 
We define this channel $\mathcal{C}$ as follows:
\begin{align}
\mathbb{P}(Y=y|X=x) =
\bordermatrix{X \backslash Y & 0 & 1 \cr
0 & 1-p & p \cr
1 & p & 1-p \cr
2 & 1-p & p \cr
3 & p & 1-p }~, \qquad
\label{eq:RBSC}
\end{align}
which is shown in Fig.~\ref{fig:RBSC_classical_factorization} (left).

To obtain the correct conditional distribution for $Y$, distinguishing $x=0$
from $x=2$ and also $x=1$ from $x=3$ are unnecessary. We eliminate this
redundancy by mapping to intermediate variable $Z$ with the function $F : \mathcal{X}\rightarrow
 \mathcal{Z}$:
\begin{align*}
F(x)=
\begin{cases}
0 \ , \ x \in \{0,2\} \\
1 \  , \ x \in \{1,3\}
\end{cases}
~.
\end{align*}
The second factor (channel $\mathcal{C}^{'}$) follows directly from this definition of $f$:
\begin{align*}
\mathbb{P}(Y=y|Z=z) =
\bordermatrix{Z \backslash Y & 0 & 1 \cr
0 & 1-p & p \cr
1 & p & 1-p}~, \qquad
\end{align*}
Figure~\ref{fig:RBSC_classical_factorization} shows this factorization for the RBSC channel. 

It is fairly easy to see that $H_\alpha\left(Z\right) \leq
H_\alpha\left(X\right)$ for any input random variable $X$ and all $\alpha\geq
0$. And so, this factorization represents a compression of the input
information, without losing information necessary for obtaining the correct
output distribution.

Now, consider the quantum map $\mathcal{E}: \mathcal{X} \to
\mathcal{B}\left(\mathcal{H}\right)$, given by:
\begin{align*}
\mathcal{E}(x) = 
\begin{cases}
\rho_A = \ketbra{A}, & \text{for } x = {0,2}\\
\rho_B = \ketbra{B}, & \text{for } x = {1,3}
\end{cases}
  ~,
\end{align*}
where:
\begin{align*}
\ket{A} &\equiv \sqrt{1-p} \ket{0} + \sqrt{p} \ket{1} ~,\\
\ket{B} &\equiv \sqrt{p} \ket{0} + \sqrt{1-p} \ket{1} ~,
\end{align*}
with orthonormal basis $\{ \ket{0}, \ket{1} \}$. If we use the measurement $\mathcal{M} =
\left\{\ketbra{0},\ketbra{1}\right\}$, we see that the channel is faithfully
represented:
\begin{align*}
\mathbb{P}(Y=0|X = x \in \{0,2\}) &= \braket{0 | \rho_A | 0} = 1-p ~,\\
\mathbb{P}(Y=1|X = x \in \{0,2\}) &= \braket{1 | \rho_A | 1}  = p ~,\\
\mathbb{P}(Y=0|X = x \in \{1,3\}) &= \braket{0 | \rho_B | 0}  = p ~,\\
\mathbb{P}(Y=1|X = x \in \{1,3\}) &=\braket{1 | \rho_B | 1}  = 1-p ~.
\end{align*}

Note that just as there were two classical intermediate states
$\mathcal{Z}=\{0,1\}$ for the RBSC, there are two quantum signal states
$\rho_0$ and $\rho_1$. For a given distribution over inputs $\mathbb{P}(X)$, we
can also compare the entropy of the classical and quantum intermediate
variables. The entropy of the quantum mixed state $\rho = \sum_{x \in \mathcal{X}}
\mathbb{P}(x) \mathcal{E}(x)$ is given by the \emph{von Neumann entropy}:
\begin{align*}
S(\rho) = - \tr(\rho \log{\rho})
  ~.
\end{align*}

\begin{figure}
\includegraphics[width=\linewidth]{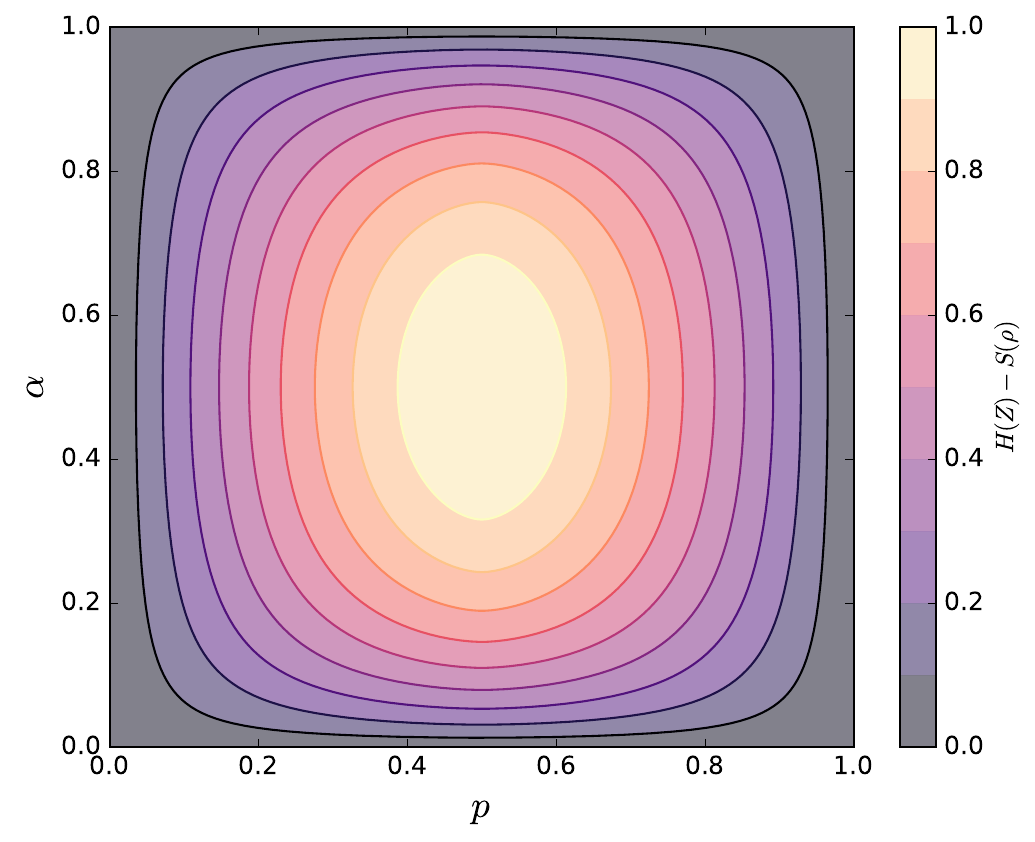}
\caption{Factorization quantum advantage: $H({Z}) - S(\rho)$ for quantum
	Redundant Binary Symmetric Channel with $\alpha = \mathbb{P}(Z=0)$; see
	Fig. \ref{fig:RBSC_classical_factorization}. Maximum advantage occurs at
	$p=\alpha=\half$ and minima when $p=0$ or $\alpha=0$.
	}
\label{fig:HEATMAP}
\end{figure}

Importantly, this entropy is less than the classical entropy:
\begin{align*}
S(\rho) \leq H({Z}) \leq H({X})
  ~,
\end{align*}
for any input distribution $\mathbb{P}(X)$ \cite{Wehr78,Hugh93}.
Figure~\ref{fig:HEATMAP} shows the \emph{quantum advantage} $H(Z) - S(\rho)$
as a function of input distribution $\mathbb{P}(X\in\{0,2\}) = \alpha$ and
channel parameter $p$. \emph{The quantum factorization of the channel is, in
this sense, more efficient than the reduced classical channel}.

\begin{figure}
\includegraphics[width=0.75\linewidth]{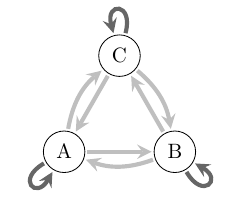}
\caption{A channel that maps the set
	$\left\{\mathrm{A},\mathrm{B},\mathrm{C}\right\}$ to itself. Dark arrows
	represent probability $p=2/3$ of transition and light arrows $p=1/6$, so
	that all the outgoing edges of a node sum to $1$.
	}
\label{fig:markov}
\end{figure}

Consider a second example. Let
$\mathcal{X}=\mathcal{Y}=\left\{\mathrm{A},\mathrm{B},\mathrm{C}\right\}$ and
consider the channel $\mathcal{C}$, depicted in Fig. \ref{fig:markov}, given by
the probabilities:
\begin{align*}
\mathbb{P}(Y=y|X=x) =
\bordermatrix{X \backslash Y & \mathrm{A} & \mathrm{B} & \mathrm{C} \cr
\mathrm{A}  & 2/3 & 1/6 & 1/6 \cr
\mathrm{B}  & 1/6 & 2/3 & 1/6 \cr
\mathrm{C}  & 1/6 & 1/6 & 2/3 }
  ~.
\end{align*}
This channel is adapted from a stochastic process in Ref. \cite{Loom18a}, and
we call it the \emph{3-MBW channel} after that process. Note that the channel
is, effectively, its own causal factorization, as each $x\in X$ has a unique
conditional distribution $\mathbb{P}\left(y\middle|x\right)$. Thus, the channel
is already its strongly advantageous factorization. With uniform input
$\mathbb{P}(x)=1/3$, we have $H_\alpha \left(X\right)=\log_2 3$ for all
$\alpha$.

Two quantum models of this channel may be constructed. The first, which we call
$\mathcal{Q}_1$, has a Hilbert space $\mathcal{H}_1$ spanned by the orthogonal
basis $\left\{\ket{\mathrm{A}},\ket{\mathrm{B}},\ket{\mathrm{C}}\right\}$ with
ensemble and POVM given by:
\begin{align*}
\mathcal{E}(x) &= \ketbra{\phi_x} ~,\\
\ket{\phi_x} &= \sum_{y\in\mathcal{Y}} \sqrt{\mathbb{P}\left(y\middle|x\right)}
\ket{y} ~,~\text{and}\\
M_y &= \ketbra{y}
  ~.
\end{align*}
Note that this induces the causal factorization. For the uniform input
$\mathbb{P}(x)=1/3$, we get an entropy cost of $S(\rho) \approx 0.69$ qubits
and a dimension cost of $S_0(\rho)=\log_2 3$ qubits. Thus, we see the model
improves on the classical case in entropy, but not in dimension. This type of
model is inspired by the \emph{$q$-machine} for processes \cite{Gu12a} and
generalizes easily to other channels.

The second model, which we call $\mathcal{Q}_0$, has a Hilbert space
$\mathcal{H}_0$ spanned by the orthogonal basis
$\left\{\ket{0},\ket{1}\right\}$ with ensemble and POVM given by:
\begin{align*}
\rho_x &= \ketbra{\psi_x} ~,\\
\ket{\psi_x} &= \begin{cases}
\ket{0} & x=\mathrm{A}\\
\frac{1}{2}\ket{0}+\frac{\sqrt{3}}{2}\ket{1} & x=\mathrm{B}\\
\frac{1}{2}\ket{0}-\frac{\sqrt{3}}{2}\ket{1} & x=\mathrm{C}
\end{cases} ~,~\text{and}\\
M_y &= \frac{2}{3}\ketbra{\psi_y}
  ~.
\end{align*}
This, too, induces the causal factorization. For the uniform input, we get an
entropy and dimension cost of $S(\rho) =S_0(\rho)= 1$ qubits. In both of these
quantifications, this model improves upon the classical case. 

The other feature worth noting is that $\mathcal{Q}_0$ is advantageous over
$\mathcal{Q}_1$ in the dimension cost $S_0$, but the opposite is true for the
entropy cost $S_1$. (Indeed, this motivated the subscripts.) Reference
\cite{Loom18a} demonstrated that, for the Markov process version of this
channel, the model $\mathcal{Q}_0$ was actually only two-dimensional model of
the process. This result trivially carries over to the channel itself, which
means $\mathcal{Q}_0$ is the model that \emph{uniquely} minimizes $S_0$.
However, it clearly does not minimize $S_1$. This demonstrates that, for at
least this channel, no ``strongly minimal'' pure-state quantum model exists.
This contrasts starkly to the causal factorization, which is strongly minimal
among classical factorizations.

%%%%%%%%%%%%%%%%%%%%
\section{Quantum Model Advantage}
\label{sec:quantum}

These examples hint at rather more general results on quantum advantage.

%%%%%%%%%%%%%%%%%%%%%%

\subsection{Single-shot: Constraints on minimal models}

Our first pair of results is motivated by the questions: What models are
optimal? And, how do they compare, generally, to the causal factorization for
classical models?

We begin with the observation, stated earlier, that each quantum model $Q$ of a
channel $\mathcal{C}$ induces a classical factorization $F_Q$. Recall that each
induced factorization must be a refinement of the causal factorization
$F_\mathcal{C}$.

Classically, a refinement to the causal factorization can be {\em merged},
where we combine causally equivalent inputs until the causal factorization is
achieved, minimizing the R{\'e}nyi entropies. Can we do the same for quantum
models? There is, in fact, a process of \emph{state-merging} that allows for
the algorithmic reduction of any R{\'e}nyi entropy individually.

The goal of state-merging is to combine the partitions of causally equivalent
states. In this case, state merging on a quantum model means modifying an
ensemble $\mathcal{E}(x)$, where $\mathcal{E}({x_1})\neq \mathcal{E}({x_2})$ for some
$x_1,x_2\in\mathcal{X}$ with
$x_1 \sim_C x_2$, into a new $\mathcal{E}'(x)$ so that
$\mathcal{E}'({x_1})= \mathcal{E}'({x_2})$. This can be accomplished in at least two ways:
\begin{enumerate}
      \setlength{\topsep}{0pt}
      \setlength{\itemsep}{0pt}
      \setlength{\parsep}{0pt}
\item $\mathcal{E}'({x_1})= \mathcal{E}({x_2})$ and $\mathcal{E}'({x})= \mathcal{E}({x})$ for $x\neq x_1$ or
\item $\mathcal{E}'({x_2})= \mathcal{E}({x_1})$ and $\mathcal{E}'({x})=
\mathcal{E}({x})$ for $x\neq x_2$.
\end{enumerate}

Recalling that the average model state is written as:
\begin{align*}
\rho = \sum_x \mathbb{P}(x) \mathcal{E}(x)
  ~,
\end{align*}
we construct two new average model states:
\begin{align*}
\rho^{(1\rightarrow 2)}
  &=  \sum_{x\neq x_1} \mathbb{P}(x)\mathcal{E}(x) + \mathbb{P}(x_1)\mathcal{E}(x_2)
  ~\text{and}\\
\rho^{(2\rightarrow 1)} 
  &=  \sum_{x\neq x_2} \mathbb{P}(x) \mathcal{E}(x) + \mathbb{P}(x_2) \mathcal{E}(x_1)
  ~.
\end{align*}
Each results from two distinct forms of state merging. It is important to
determine if such a technique reduces memory costs. This leads to a lemma that
underlies our first main result.

\begin{Lem}
Let $f:\mathbb{R}\rightarrow\mathbb{R}$ be a concave function and let $S_f(\rho) = \mathrm{Tr}\left(f(\rho)\right)$. Then:
\begin{align}
\label{eq:bound}
S_f(\rho)
\geq \min\{S_f(\rho^{(1\rightarrow 2)}), S_f(\rho^{(2\rightarrow 1)})\}
  ~.
\end{align}
\end{Lem}
\begin{ProLem}
Let $f:\mathbb{R}\rightarrow\mathbb{R}$ be a concave function and
let $S_f(\rho) = \mathrm{Tr}\left(f(\rho)\right)$. Define:
\begin{align*}
\alpha &=\mathbb{P}(x_1)/ (\mathbb{P}(x_1)+\mathbb{P}(x_2)) \\
1-\alpha &= \mathbb{P}(x_2)/ (\mathbb{P}(x_1)+\mathbb{P}(x_2))
  ~.
\end{align*}
One can check that:
\begin{align*}
\rho = \alpha \rho^{(1\rightarrow 2)} + (1-\alpha) \rho^{(2\rightarrow 1)}
  ~.
\end{align*}
By $f$'s concavity:
\begin{align}
S_f(\rho) &\geq \alpha S_f(\rho^{(1\rightarrow 2)})
  + (1-\alpha) S_f(\rho^{(2\rightarrow 1)}) \nonumber \\
  & \geq \min\{S_f(\rho^{(1\rightarrow 2)}), S_f(\rho^{(2\rightarrow 1)})\}
  ~.
\end{align}
So, either $\rho^{(1\rightarrow 2)}$ or $\rho^{(2\rightarrow 1)}$ must be
smaller according to $f$.
\end{ProLem}

From this useful bound follows one of our primary results: 

\begin{Result}[Causal Model Optimality]
A quantum model of a channel $\mathcal{C}$ minimizes $S_f\left(\rho\right)$, where $\rho$ is the average of the models' ensemble and $f$ is any concave function, \emph{only if} it induces the causal factorization of $\mathcal{C}$.
\end{Result}

To see this, suppose otherwise: There exists a quantum model $\mathcal{Q}$ that
does not induce the causal factorization, but that is $f$-minimal. By Eq.
\eqref{eq:bound}, though, we can perform a state-merging that yields a model
$\mathcal{Q}'$ with $S_f(\rho') \leq S_f(\rho)$. This is a contradiction.

Of course, this result applies to the R{\'e}nyi entropies
$S_\alpha\left(\rho\right)$ as well:
$S_\alpha\left(\rho\right)=S_f(\rho)/(1-\alpha)$, where $f(x) = x^\alpha$.

There are some caveats to this result. The first is that a minimum does not
\emph{necessarily} exist. The space of all possible models contains models of
arbitrarily high Hilbert space dimension, and so it is not compact. This is
still true even if we restrict to the space of models that induce the causal
factorization: There is no restriction on the dimension of the states
$\mathcal{E}(x)$.

However, if we restrict to pure-state models that induce the causal
factorization, then the states $\mathcal{E}(x)$ can span a space of dimension
no larger than $\dim\mathcal{H} = \left|\mathcal{Z}_\mathcal{C}\right|$. This
makes the model-space compact and ensures the existence of a minimal model for
each $S_\alpha$.

Now, in the classical case, the causal factorization was unique and {\em
strongly minimal}, in that it minimized all $H_\alpha$. Our result shows that
only quantum models that induce the causal factorization can be minimal.
However, it does not prove that \emph{strongly} minimal models exist. As we saw
in the example of the 3-MBW channel, there may be models that are only
\emph{weakly} minimal, in that they only minimize some R{\'e}nyi entropies and
not others.

Our second main result focuses on comparing pure-state quantum models with the
classical causal factorization itself. Pure-state quantum models, recall, are a
quantum analogue to the factorization. Furthermore, as we just saw, pure-state
quantum models that minimize any R{\'e}nyi entropy must induce the causal
factorization. Thus, to determine if and when they improve upon their classical
counterpart, it makes sense to compare pure-state models that induce the causal
factorization with the causal factorization itself.

Let $\mathcal{Q}$ be a pure-state quantum model that induces the causal
factorization. Then its average model state is:
\begin{align*}
\rho = \sum_{z} \mathbb{P}(z) \ket{\psi_z}\bra{\psi_z}
  ~,
\end{align*}
where the sum is over causal states $z\in\mathcal{Z}_\mathcal{C}$ and the
states $\ket{\psi_z}$ correspond to the states of the quantum model:
$\mathcal{E}(x) = \ketbra{\psi_z}$, if $z = f_\mathcal{C}(x)$. 

We can also consider the orthogonal decomposition:
\begin{align*}
\rho = \sum_{i} \lambda_i \ket{i}\bra{i}
  ~.
\end{align*}
The vector of coefficients $\lambda_i$ and the probability distribution
$\mathbb{P}(z)$ are related by a unitary matrix $U_{iz}$ such that
\cite{Hugh93}:
\begin{align*}
\mathbb{P}(z) = \sum_{i} \left|U_{iz}\right|^2 \lambda_i
  ~.
\end{align*}
$U_{iz}$ is the identity matrix if and only if the states
$\left\{\ket{\psi_z}\right\}$ are orthogonal. This establishes a
\emph{majorization} relationship between the two vectors, written $\lambda_i
\succsim \mathbb{P}(z)$ \cite{Mars11a}. The importance of this result here is
that, for all concave $\alpha\geq 0$, we have $S_\alpha(\rho) \leq
H_\alpha\left(Z_{\mathcal{C}}\right)$; with equality for $\alpha>0$ if and only
if the states $\left\{\ket{\psi_z}\right\}$ are orthogonal.

This gives our second main result:
\begin{Result}[Strong/strict advantage of quantum models]
Given any input distribution $\mathbb{P}(x)$, a pure-state quantum model of a
channel $\mathcal{C}$ that induces the causal factorization $F_\mathcal{C}$
satisfies $S_\alpha(\rho) \leq H_\alpha\left(Z_{\mathcal{C}}\right)$, where
$\rho := \sum_x \mathbb{P}(x)\mathcal{E}(x)$ is the average state. For
$\alpha\geq 0$, the above bound is strict whenever the ensemble
$\mathcal{E}(x)$ is nonorthogonal.
\end{Result}
This result generalizes a similar result from Ref. \cite{Loom18a} to channels.
In fact, it strengthens it. Rather than $\mathbb{P}(z)$ being determined by a
process' unique stationary state distribution, here it can be determined by any
input distribution $\mathbb{P}(x)$, and the quantum model state is still
strongly advantageous.

%%%%%%%%%%%%%%%%%%%%
\subsection{Asymptotics: A lower bound}
\label{sec:asymptotic}

The previous section focused on pure states. Now, we return to mixed states to
study the general asymptotic case.

Earlier we defined the $(n,M,\epsilon)$-quantum model. We introduce here a
single-letter optimization that lower-bounds the rate $\frac{1}{n}\log M$ at
which a channel can be quantally modeled, based on the classical Wyner common
information. Unlike the classical case, we do not claim this bound is tight.

To set it up, consider a $d$-dimensional model
$\left(\mathcal{H},\mathcal{E},\mathcal{M}\right)$ of the variables $XY$, with
$\rho=\sum_x \mathbb{P}(x)\mathcal{E}(x)$. We define the \emph{Holevo information}:
\begin{align*}
I_{\chi}\left(X:\rho\right) = S(\rho) - \sum_x  \mathbb{P}(x) S\left(\mathcal{E}(x)\right)
  ~.
\end{align*}
For each state in the ensemble, we do an orthogonal decomposition:
\begin{align*}
\mathcal{E}(x) = \sum_{i=1}^d  \lambda_{i|x} \ketbra{i,x}
  ~,
\end{align*}
and define the conditional measurement information:
\begin{align*}
I_\mu&\left(Y:\rho\middle|X\right)\\
&=\sum_{x\in\mathcal{X}} \sum_{i=1}^d \mathbb{P}\left(x,i,y \right)
\log\left(\frac{\mathbb{P}\left(i,y \middle| x\right)}{\mathbb{P}\left(y \middle| x\right)\lambda_{i|x}}\right)
\end{align*}
where:
\begin{align*}
\mathbb{P}\left({x, i, y} \right) = \mathbb{P}(x) \lambda_{i|x} \braket{i,x | M_y |i,x}
  ~.
\end{align*}
(Note that $\mathbb{P}({x,y}) = \sum_{i} \mathbb{P}\left({x, i,y} \right)  $.) With these two, we define:
\begin{align*}
I\left(XY:\rho\right)=I_{\chi}\left(X:\rho\right)+I_\mu\left(Y:\rho\middle|X\right)
  ~.
\end{align*}
Another way to approach this quantity is to define:
\begin{align*}
\rho_{xy} =  \sum_{i=1}^d \frac{\lambda_{i|x}\braket{i,x | M_y |i,x}}{\mathbb{P}\left(y\middle|x\right)} \ket{i,x}\bra{i,x}
\end{align*}
as a state representing our knowledge, after the measurement, about what $i$
\emph{was} before measurement. (This is not the final state after measurement,
as the measurement itself induces further modification.) We then have:
\begin{align*}
I\left(XY:\rho\right)=S(\rho) - \sum_{x,y} \mathbb{P}\left(x,y\right) S\left(\rho_{xy}\right)
  ~.
\end{align*}
This is equivalent to our earlier definition, but expresses the quantity as a
kind of Holevo quantity.

We define an analogue to the Wyner common information as:
\begin{align*}
C_{q,l}\left(X\middle>Y\right)= \inf_{X-\rho-Y} I\left(XY:\rho\right)
  ~.
\end{align*}
where $X-\rho-Y$ indicates that we are varying over all models of $XY$. The notation $\left.X\right>Y$ indicates that this is not a symmetric quantity
between $X$ and $Y$.

Notice that we define this quantity as an infimum and not a minimum, as Wyner
did. This is because Wyner's proof \cite{Wyne75a} demonstrated that the infimum
could be attained on a compact subset of the model space. This result relies on
being able to decompose $I\left(XY:W\right)$ as $H\left(XY\right) -
H\left(XY\middle|W\right)$. The minimization of $I\left(XY:W\right)$ becomes
the maximization of $H\left(XY\middle|W\right)$. And this can be written as a
linear function of $\mathbb{P}(w)$:
\begin{align*}
H\left(XY\middle|W\right) = -\sum_w \mathbb{P}(w)\sum_{x,y} \mathbb{P}\left(x,y\middle|w\right)
\log\mathbb{P}\left(x,y\middle|w\right)
  ~,
\end{align*}
if we hold $\mathbb{P}\left(x,y\middle|w\right)$ fixed. 

It is the linearity of the optimization function, in the classical case, that
allows the assumption of a compact minimization. However, our quantum
generalization $I\left(XY:\rho\right)$ cannot be expressed in a way that allows
the optimization to be linearized. This is why we cannot assume the existence
of a minimum.

Our third main result can now be stated as follows.
\begin{The}
Let $\delta>0$. Then there exists a fixed $\Delta>0$ such that, for all quantum
$(n,M,\epsilon)$-models, $\frac{1}{n}\log M \leq C_{q,l}\left(X:Y\right) -
\delta$ implies $\epsilon > \Delta$. 
\end{The}
In other words, if the model's cost rate is bounded below by $C_{q,l}$, then
its error is bounded away from zero. To achieve arbitrary accuracy, it must
have a rate of at least $C_{q,l}$.

\begin{ProThe}
The argument follows the same logic as Wyner's \cite{Wyne75a} and proceeds by
contradiction. Suppose with $\delta$ fixed that, for any $\Delta>0$, we can
find a $(n,M,\epsilon)$-model with $\frac{1}{n}\log M \leq
C_{q,l}\left(X:Y\right) - \delta$ and $\epsilon < \Delta$. For such a model:
\begin{align*}
\frac{1}{n} \log M &\geq \frac{1}{n}I\left(X^n \widehat{Y}^n:\rho\right)\\
&\geq \frac{1}{n} H\left(X^n \widehat{Y}^n\right) - \frac{1}{n}H\left(X^n
\widehat{Y}^n\middle| \rho\right)
  ~,
\end{align*}
where:
\begin{align*}
H\left(X^n \widehat{Y}^n\middle| \rho\right) &= H_\chi\left(X^n\middle|\rho\right) + H_\mu\left(Y^n\middle|\rho, X^n\right)
  ~,\\
H_\chi\left(X^n\middle| \rho\right) &= H(X^n) - I_\chi\left(X^n:\rho\right)
  ~,~\text{and}\\
H_\mu \left( \widehat{Y}^n\middle| \rho,X\right) &= H\left(\widehat{Y}^n\right) - I_\mu\left(Y^n:\rho\middle|X^n\right)
  ~.
\end{align*}
We see that $H_\mu \left( \widehat{Y}^n\middle| \rho,X\right)$ is explicitly a
conditional entropy by definition and so is subadditive. The subadditivity of
$H_\chi\left(X^n\middle| \rho\right)$ is follows from the strong subadditivity
of von Neumann entropy. Subadditivity implies that:
\begin{align*}
H\left(X^n \widehat{Y}^n\middle| \rho\right)\leq \sum_{k} H\left(X_k \hat{Y}_k\middle|\rho\right)
  ~.
\end{align*}
We then have:
\begin{align*}
\frac{1}{n} \log M 
&\geq \frac{1}{n} H\left(X^n \widehat{Y}^n\right) -
\frac{1}{n}\sum_{k}H\left(X_k \widehat{Y}_k\middle| \rho\right)
  ~.
\end{align*}
By assumption, we can drive $\Delta$ low enough that $H\left(X^n
\widehat{Y}^n\right)\geq H(X^n Y^n) - O\left(\Delta\right)$ where $O\left(\Delta\right)$
is the standard big-$O$ notation. Since
$H\left(X^n Y^n\right)$ is separable, we can write $H\left(X^n
\widehat{Y}^n\right) = \sum_k H\left(X_k \widehat{Y}_k\right) - O\left(\Delta\right)$ for some
$\eta>0$. Thus:
\begin{align*}
\frac{1}{n} \log M 
&\geq \frac{1}{n} \sum_k I\left( X_k \widehat{Y}_k : \rho \right) - O\left(\Delta\right)
  ~.
\end{align*}
This is a sum over a collection of single-letter models of channels that are at
least $\Delta$-close to $XY$. Each such model is then $\Delta$-close to another
model $\rho_k$ that perfectly generates $XY$. In each case we have $I\left( X_k
\widehat{Y}_k : \rho \right) \geq I\left( X Y: \rho_k
\right)-O\left(\Delta\right)$. And, for each $k$ we have $I\left( X Y: \rho_k
\right)\geq C_{q,l}\left(X\middle>Y\right)$. Then:
\begin{align*}
\frac{1}{n} \log M
  & \geq \frac{1}{n} \sum_k C_{q,l}\left(X\middle>Y\right) -
  O\left(\Delta\right) \\
  & = C_{q,l}\left(X\middle>Y\right)-\frac{1}{n}O\left(\Delta/n\right)
  ~.
\end{align*}
Now, take $\Delta$ small enough so that the error term is less than $n\delta$.
We get:
\begin{align*}
\frac{1}{n} \log M 
&\geq C_{q,l}\left(X\middle>Y\right)-\delta
  ~.
\end{align*}
However, this contradicts the assumption that $\frac{1}{n} \log M \leq
C_{q,l}\left(X\middle>Y\right)-\delta$, completing the proof.
\end{ProThe}

Thus, $C_{q,l}\left(X\middle>Y\right)$ offers a lower-bound on the rate at
which one can model the channel $\mathcal{C}$. We now discuss the ways in which
this lower-bound is lacking.

First, as we mentioned before, there is no guarantee that this bound can be
easily computed via an optimization over a compact set. This arises from the
optimization's nonlinearity.

Second, it lacks a direct counterpart---that is, a theorem showing that
$C_{q,l}\left(X\middle>Y\right)$ is achievable and not just a lower bound.
Again, this stems from a fundamental difference between the setting of Wyner's
classical proof and the quantum models we are considering. Wyner's
achievability proof turns on showing that, given a single-shot model of a
channel $\mathcal{C}$, one can take asymptotically many copies of the
single-shot model and use random coding to trim off excess information.

For a quantum model, though, random coding means the random selection of
subspaces. If the states $\rho_x$ of the ensemble commute, we may choose
subspaces that commute with them all. Otherwise, coherence effects destroy the
random coding approximation. In other words, there is a trade-off between
minimizing the size of the single-shot model through the use of quantum
coherence and, if coherence is present, the inability to further reduce the
rate using asymptotic random coding. Finding an achievable lower bound on
quantum models requires balancing these two tensions.

Finally, we note simply that $C_{q,l}\left(X\middle>Y\right)$ is not symmetric.
While this may not seem necessary---after all, the channel simulation question
is not one that is manifestly symmetric---we show in the following section that
the quantum simulation of channels is equivalent to another question, involving
entanglement, that \emph{is} manifestly symmetric. We propose that this
alternative formulation offers a more productive avenue of investigation.

%%%%%%%%%%%%%%%%%%%%%%
\subsection{Equivalence to common entanglement}
\label{sec:entanglement}

Suppose we have a channel $\mathcal{C}:\mathcal{X}\rightarrow \mathcal{Y}$, a
quantum model $\mathcal{Q}=\left(\mathcal{H},\mathcal{E},\mathcal{D}\right)$,
and an input $X\sim \mathbb{P}(x)$. We will show how to reformulate these
elements into a symmetric picture, where the joint random variable $XY$ is
generated by taking local measurements on a bipartite entangled state
$\ket{\Psi_{AB}}$. Just as the generation of a joint random variable from a
classical shared variable with local operations is called common
information, the generation of a joint random variable from shared
entanglement is here called \emph{common entanglement}.

For a mixed state $\rho$ on a space $\mathcal{H}_B$
with dimension $d$, we can write
its orthogonal decomposition as:
\begin{align*}
\rho = \sum_{i=1}^d \lambda_i \ket{i}_B\ket{i}_B
  ~.
\end{align*}
We assume without loss of generality that $\mathcal{H}_B$ is entirely spanned
by $\rho$'s support; that is, $\lambda_i>0$ for all $i=1,\dots,d$. We define
$\rho$'s \emph{purification} as a state that exists on the space
$\mathcal{H}_A\otimes \mathcal{H}_B$, where $\mathcal{H}_A$ is a copy of
$\mathcal{H}_B$ with dimension $d$ as well. $\rho$'s purification is:
\begin{align*}
\ket{\psi}_{AB} = \sum_{i=1}^d \sqrt{\lambda_i} \ket{i}_A\otimes \ket{i}_B
  ~,
\end{align*}
where the basis $\{\ket{i}_A\}$ is unitarily equivalent to the original
$\{\ket{i}_B\}$. It is a purification in the sense that
$\mathrm{Tr}_B\left(\ket{\psi}_{AB}\right)=\rho$. That is, one can imagine
the mixed state only exists due to ignorance of a larger system. Up to
unitary transformations on $\mathcal{H}_A$, this purification is unique.

Ensembles $\mathcal{E}:\mathcal{X}\rightarrow\mathcal{B}_{+}\left(\mathcal{H}\right)$
may also be purified. Our scheme here resembles that in Ref.
\cite{Bass03a} except that we allow the ensemble states $\mathcal{E}\left(x\right)$ to be mixed.
Given an input $\mathbb{P}\left(x\right)$, we write:
\begin{align*}
\rho & = \sum_{x\in\mathcal{X}} \mathcal{E}(x) \\
     & = \sum_{x\in\mathcal{X}} \mathbb{P}(x) \rho_x
\end{align*}
as the ensemble decomposition of $\rho$ on $\mathcal{H}_B$. Let
$\ket{\psi}_{AB}$ be a purification of $\rho$ and $\ket{\psi_x}_{AB}$ be a
purification of the {\em unnormalized} state $ \mathbb{P}(x)\mathcal{E}(x)$ (both onto
$\mathcal{H}_{A}\otimes \mathcal{H}_B$). 

Letting $\left\{\ket{i,x}\right\}$ be the diagonalizing basis of
$\mathbb{P}(x)\mathcal{E}(x)$ and $\lambda_{i|x}$ be its spectra, we can write:
\begin{align*}
\ket{\psi_x}_{AB} & = \sum_{i=1}^d \sqrt{\lambda_{i|x}} \ket{i,x}_A \otimes\ket{i,x}_B\\
& = \sum_{j=1}^d\sum_{i=1}^d \sqrt{\lambda_{i|x}}
\ket{i,x}_A\otimes \left(\braket{j | i,x} \ket{j}_B\right) \\
& = \sum_{j=1}^d \sqrt{\lambda_j}
\left(\sum_{i=1}^d \sqrt{\frac{\lambda_{i|x}}{\lambda_{i|x}}}
\braket{j | i,x} \ket{i,x}_A\right) \otimes\ket{j}_B \\
& = \sum_{\omega\in\Omega} \sqrt{\lambda_j} \ket{\Phi_{j,x}}_A \otimes \ket{j}_B
  ~,
\end{align*}
where:
\begin{align*}
\ket{\Phi_{j,x}}_A = \sum_{i=1}^d \sqrt{\frac{\lambda_{i|x}}{\lambda_{j}}}
\braket{j | i,x} \ket{i,x}_A
\end{align*}
For each $x\in\mathcal{X}$, there is a unique operator $K_x$ on $\mathcal{H}_A$
such that:
\begin{align*}
K_x\ket{j}_A = \ket{\Phi_{j,x}}_A
  ~.
\end{align*}
Furthermore:
\begin{align*}
\braket{j | \sum_x K^{\dagger}_xK_x | j'}
  & =\sum_{x\in\mathcal{X}} \braket{\Phi_{j,x} | \Phi_{j',x}}\\
  & = \sum_{x\in\mathcal{X}} \sum_{i=1}^d
  \frac{\lambda_{i|x}}{\sqrt{\lambda_j \lambda_{j'}}}
  \braket{j | i,x} \braket{i,x| j '}\\
  & = \frac{1}{\sqrt{\lambda_j \lambda_{j'}}}
  \braket{j | \rho| j '} = \delta_{jj'}
  ~.
\end{align*}
This shows that the $\left\{K_x\right\}$ are the \emph{Kraus operators} of a
complete POVM $\mathcal{M}=\left\{M_x\right\}$ with $M_x = K_x^\dagger K_x$.

To summarize, an ensemble $\mathcal{E}(x)$ has a unique purification
$\left(\ket{\psi}_{AB},\mathcal{M}_A\right)$ satisfying:
\begin{align*}
\mathrm{Tr}\left(\ketbra{\psi}_{AB}\right) &= \rho_B\\
\mathrm{Tr}\left(M_x \ketbra{\psi}_{AB}\right) &= \mathbb{P}\left(x\right) \mathcal{E}\left(x\right)
  ~,
\end{align*}
where $\rho_B=\sum_x \mathbb{P}(x)\mathcal{E}(x)$.
This purification is unique up to transformations of the form $K_x\mapsto U K_x
V^{\dagger}_x$ for unitary matrices $U$ and $\left\{V_x\right\}$ on
$\mathcal{H}_A$. (These arise from the unitary ambiguity of each individual
purification of $\rho$ and $\mathcal{E}(x)$.)

Correspondingly, the \emph{common entanglement representation} of a quantum
model $\mathcal{Q}=\{\mathcal{H}_B,\mathcal{E},\mathcal{M}\}$ is the triplet
$\Phi=\left(\ket{\Phi}_{AB}, \mathcal{M}_A,\mathcal{M}_B\right)$, where
$\left(\ket{\Phi}_{AB},\mathcal{M}_A\right)$ is the purification of the
ensemble $\mathcal{E}$ and $\mathcal{M}_B = \mathcal{M}$.

\begin{figure}
\includegraphics[width=0.75\linewidth]{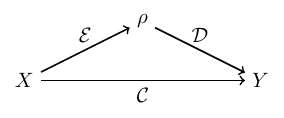}
\includegraphics[width=0.75\linewidth]{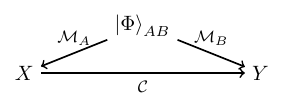}
\caption{Two equivalent pictures: (Top) Directional picture modeling a channel
	$\mathcal{C}:\mathcal{X}\rightsquigarrow\mathcal{Y}$ with an intermediate
	quantum system $\rho$. (Bottom) Symmetric picture generating a joint random
	variable $XY$ from an entangled state $\ket{\Phi}_{AB}$.
}
\label{fig:quant_diagram}
\end{figure}

In this representation the distribution $\mathbb{P}\left(x,y\right)$ is seen as
being generated by the local POVM $\mathcal{M}_A\otimes \mathcal{M}_B =
\left\{M_x\otimes N_y\right\}$ from a shared entangled pair. In the language of
entanglement theory the representation is equivalent to simulating the channel
via a quantum ensemble. See Fig. \ref{fig:quant_diagram}.

For every quantum model $\mathcal{Q}=\{\mathcal{H}_B,\mathcal{E},\mathcal{M}\}$
of the channel $\mathcal{C}:\mathcal{X}\rightsquigarrow \mathcal{Y}$ and input
random variable $X\sim \mathbb{P}\left(x\right)$, the joint random variable
$XY$ has a common entanglement generator $\Phi=\left(\ket{\Phi}_{AB},
\mathcal{M}_A,\mathcal{M}_B\right)$. This alternative perspective on the
reverse Holevo problem will prove fruitful for future extensions of our
results.

\section{Concluding remarks}
\label{sec:conclusion}

We inverted the question originally posed by Holevo. Rather than using classical
resources to quantify the behavior of quantum channels, we analyzed how well
quantum resources may simulate the workings of a classical channel.

This question makes sense in the context of wider considerations. We do
not claim that it is practically efficient to replace classical channels with
quantum ones. Attempts to construct and use quantum channels, such as in
quantum teleportation, often require classical communication to work. However,
this is a consequence of the difficulties in transmitting quantum information
over distances. When we discuss channels as a microcosm of the inner workings
of a stochastic process, in contrast, we are invoking information that is
retained over time. This information is memory. When we measure the R{\'e}nyi
entropies of the average quantum state of a model, we are subjecting it to
various memory metrics.

Thus, our results here are best interpreted as a study not in spatial
communication, but in the memory required to simulate a temporal channel. In
this setting, we showed that:
\begin{enumerate}
      \setlength{\topsep}{-10pt}
      \setlength{\itemsep}{-3pt}
      \setlength{\parsep}{-10pt}
\item In the single-shot regime, those quantum models minimizing various
	memory metrics are those that induce the causal state.
\item Conversely, pure-state models that induce the causal state improve upon
	the best classical factorization across all memory metrics, and
\item In the asymptotic regime, for arbitrary accuracy to be achieved the
	quantum memory rate required for simulation is bounded below by the new
	quantum common information quantity for
	channels---$C_{q,l}\left(X\middle>Y\right)$---introduced here.
\end{enumerate}
Many questions remain. For instance, under what conditions is
$C_{q,l}\left(X\middle>Y\right)$ achievable? When it is not, can we derive a
tighter lower bound? The common entanglement representation of the problem
in Section \ref{sec:entanglement} should prove useful in furthering study
of these questions.

It also remains to rigorously connect our results for channels back to the
problem of simulating stochastic processes with quantum resources. The rates
achievable for channels can be applied to the \emph{process channel}
$\mathbb{P}\left(x_{0:\infty}\middle|x_{-\infty:0}\right)$ from the past to the
future. These would give lower bounds on the rates at which we can
\emph{sequentially generate} a process. Likely, they would not be tight. Thus,
as results for simulating channels are derived, the additional constraints
characterizing sequential generation must also be formulated.

% \section*{Author Contributions}
% 
% All authors researched, collated, and wrote this article.
% 
% \section*{Competing interests}
% 
% The authors declare no conflicts of interest.

\vspace{-0.2in}
\section*{Acknowledgments}
\label{sec:acknowledgments}
\vspace{-0.2in}

We thank Ryan James and Fabio Anza for useful conversations. As a faculty
member, JPC thanks the Santa Fe Institute and the Telluride Science Research
Center for their hospitality during visits. This material is based upon work
supported by, or in part by, the John Templeton Foundation grant 52095,
Foundational Questions Institute grant FQXi-RFP-1609, and U. S. Army Research
Laboratory and the U. S. Army Research Office under contract W911NF-13-1-0390
and grant W911NF-18-1-0028.

\bibliography{chaos,ref,Qchannel,maj-ref}

\end{document}